\documentclass[prd,amsmath,
twocolumn,floatfix,amssymb, preprintnumbers, nofootinbib, superscriptaddress]{revtex4} 

\usepackage{soul}
\usepackage{amsmath,amssymb,amsfonts}
\usepackage{graphicx}
\usepackage{hyperref}
\usepackage{slashed}
\usepackage{booktabs,tabulary}
\usepackage{mciteplus}
\usepackage{color}
\usepackage{esint}
\usepackage{calrsfs}
\usepackage{nicefrac}
\usepackage{physics}
\usepackage[percent]{overpic}

\usepackage{bibentry}
\newcommand{\ignore}[1]{}
\newcommand{\nobibentry}[1]{{\let\nocite\ignore\bibentry{#1}}}

\makeatletter
\def\bibinfo@X@title#1,{\ignorespaces}
\makeatother

\def\barr{\left(\begin{array}{c}}
\def\earr{\end{array}\right)}
\def\bmat{\left(\begin{array}{cc}}
\def\emat{\end{array}\right)}

\def\beq{\begin{equation}}
\def\eeq{\end{equation}}
\def\bea{\begin{eqnarray}}
\def\eea{\end{eqnarray}}
\def\beqa{\begin{equation}\begin{array}{l}}
\def\eeqa{\end{array}\end{equation}}

\def\dd{{\rm d}}

\renewcommand{\arraystretch}{1.5}

\begin{document}

\title{Light-by-light scattering sum rule for radiative transitions
of bottomonia}
\author{Victor Ananyev}\email{victor.ananyev@gmail.com}
\affiliation{Institut f\"ur Kernphysik \& PRISMA$^+$  Cluster of Excellence, Johannes Gutenberg Universit\"at,  D-55099 Mainz, Germany}
\affiliation{Department of Physics, Taras Shevchenko National University of Kyiv, 6 Academician Glushkov Avenue., Kyiv 03680, Ukraine}
\author{Igor Danilkin}\email{danilkin@uni-mainz.de}
\affiliation{Institut f\"ur Kernphysik \& PRISMA$^+$  Cluster of Excellence, Johannes Gutenberg Universit\"at,  D-55099 Mainz, Germany}
\author{Marc Vanderhaeghen}\email{vandma00@uni-mainz.de}
\affiliation{Institut f\"ur Kernphysik \& PRISMA$^+$  Cluster of Excellence, Johannes Gutenberg Universit\"at,  D-55099 Mainz, Germany}

\date{\today}

\begin{abstract}
We generalize a forward light-by-light scattering sum rule to the case of heavy quarkonium radiative transitions. We apply such sum rule to the bottomonium states, and use available data on radiative transitions in its evaluation. For the transitions that are not known experimentally, we provide theoretical estimates within a potential model, and consider the spread between similar  approaches in the literature as an estimate for the model error. For the $\Upsilon(1S)$, $\Upsilon(2S)$, and $\Upsilon(3S)$ states we observe that, due to a cancellation between transitions involving $\chi_{b0}, \chi_{b1}$, and $\chi_{b2}$ states, the sum rule is satisfied within experimental and theoretical error estimates. Having tested this sum rule for the low-lying bottomonium states, it may be used as a tool to investigate the nature of exotic states in the charmonium and bottomonium spectrum through the corresponding radiative transitions. 
\end{abstract}

\maketitle

\section{Introduction}
Several model-independent sum rules were derived  for the forward light-by-light scattering, and were  exactly verified at leading order in scalar and spinor QED~\cite{Gerasimov:1965et,Drell:1966jv,Drechsel:2004ki,Pascalutsa:2010sj,Pascalutsa:2012pr}. Such sum rules are valid for the case when at least one photon is real and the other is spacelike or below the first particle production threshold, i.e., for photon virtualities $ q_1^2 \leq s_0$, $q_2^2 = 0$, where $s_0$ is the particle production threshold. Three of these sum rules have the form of a superconvergence relation, for which an integral over an experimentally measurable quantity yields  zero~\cite{Pascalutsa:2012pr}. One of these is a helicity sum rule of the form
\begin{align}
 \int_{s_0}^{\infty}\frac{ds}{(s-q_1^2)}\,
 \left(\sigma_2 - \sigma_0\right)_{q_2^2=0}=0,\label{Eq:SumRuleV1}
\end{align}
where $\sigma_0$ and $\sigma_2$ are the total helicity cross sections for the $\gamma^\ast \gamma \to X$ processes for total helicity 0 and 2,  respectively, where $X$ denotes the sum over all allowed final states. Such light-by-light sum rules have been applied within different field theories in both perturbative and nonperturbative settings~\cite{Pascalutsa:2010sj,Pascalutsa:2012pr,Pauk:2013hxa}. Furthermore, their application to the $\gamma^\ast \gamma$ production of light-quark mesons has been discussed in Refs.~\cite{Pascalutsa:2012pr,Danilkin:2016hnh}, and the application to the $\gamma \gamma$ production of charmonium states was discussed in Ref.~\cite{Danilkin:2017utg}. For the  pseudoscalar, scalar, axial-vector, and tensor mesons, where $\gamma^\ast \gamma \to X$ data are available, these sum rules were verified within 10\% - 30\% experimental accuracy~\cite{Danilkin:2016hnh}.

In the present work we investigate the extension of such sum rules, when one of the virtual photons is replaced by a vector quarkonium state. For the conventional heavy-quark $Q \bar Q$ bound states, radiative transitions have been measured quite extensively in the past decades by collaborations at the charm and $B$ factories, CLEO@CESR,   BABAR@PEP-II, Belle@KEKB, and BESIII@BEPCII, and will be studied in the near future by Belle-II.  

The study of light-by-light sum rules in the heavy quarkonium sector may also be worthwhile in light of the plethora of new states, so-called $XYZ$ states, which have been found in recent years above open heavy flavor thresholds at all of these facilities, see e.g. Refs.~\cite{Chen:2016qju,Lebed:2016hpi,Olsen:2017bmm} for some recent reviews and references therein. Such sum rule relations have the potential to reveal how much of the radiative decay strength from or into vector quarkonium states results from possible exotic mesons. An  example is the $X(3872)$ state with $J^{PC} = 1^{++}$, which sits right at the $D \bar D^\ast$ threshold, for which the radiative transitions $\gamma J/\psi$ and $\gamma \psi^\prime$ have been proposed as a diagnostic tool for studying the nature of this state~\cite{Swanson:2004pp}, shedding light on its hybrid  charmonium-molecular nature. Rare decays of $X(3872)$ will be an important part of the PANDA~\cite{Lutz:2009ff} scientific program where such studies are feasible even at the start of data taking. Also, at BESIII the first radiative transition between two exotic mesons has been observed in the process $Y(4260) \to \gamma X(3872)$~\cite{Ablikim:2013dyn}, and detailed studies of radiative transitions can be expected from Belle-II in the near future~\cite{Kou:2018nap}.

The outline of this work is as follows. In Section~\ref{sec:introsr}, we introduce the helicity sum rule that we will study in this work for radiative transitions between quarkonium states, of which one has $J^{PC} = 1^{--}$ quantum numbers. In Section~\ref{sec:spectr} we describe the potential model adopted from Refs.~\cite{Deng:2016stx,Deng:2016ktl} to reproduce heavy quarkonium wave functions. In Section~\ref{sec:rad-trans} we review the formalism to evaluate the leading radiative transitions between quarkonium states with defined total helicity, and make a comparison between available experimental values and theoretical results in the literature.  In Section~\ref{sec:sr} we make use of experimental information on the radiative transitions  $\Upsilon(mS) \to \gamma \chi_{bJ}(nP)$ for $m>n$, as well as theoretical estimates for  $\chi_{bJ}(nP) \to \gamma \Upsilon(mS)$ for $n \geq m$, and evaluate the derived helicity sum rule. We provide a quantitative discussion for the $\Upsilon(1S)$, $\Upsilon(2S)$, and $\Upsilon(3S)$ states. Finally, a summary and outlook is given in Section~\ref{sec:conclude}.

\section{Sum rule for quarkonium radiative transitions}
\label{sec:introsr}

\begin{figure}
\includegraphics[width=0.9\linewidth]{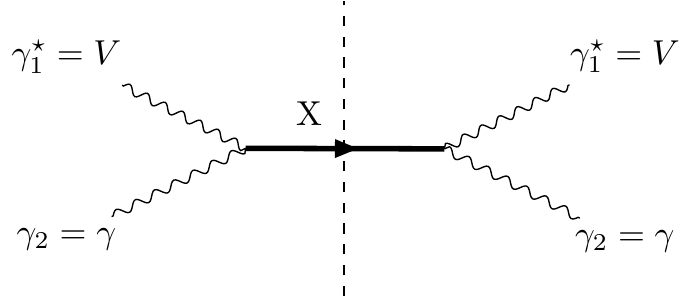}
\caption{Light-by-light forward scattering involving one virtual ($V$) and one on-shell $(\gamma)$ photon. We associate $V$ with a vector quarkonium state. As a result of the optical theorem (dashed cut), intermediate quarkonium states $X$ with $J^{PC} = 0^{-+}, 0^{++}, 1^{++}, 2^{++}$, ... contribute to the forward scattering.}
\label{fig:lbl-scatt-cut}
\end{figure}

In the case when one photon in Eq.~(\ref{Eq:SumRuleV1}) is replaced by a vector quarkonium state, one needs to take into account its nonpointlike structure. In full analogy to the original Gerasimov-Drell-Hearn sum rule~\cite{Gerasimov:1965et,Drell:1966jv,Drechsel:2004ki}, it gives the relation to the anomalous magnetic moment $a_V$. Therefore, for the states below the $B\bar{B}$ threshold ($m_V^2 < 4\,m_B^2$) it holds that
\begin{align}
&\text{SR}\equiv \int_{0}^{\infty}\frac{ds}{(s-m_V^2)^2}\,
\left(\text{Im}\,M_{+-,+-} - \text{Im}\,M_{++,++}\right)_{q_2^2=0}
\nonumber \\
&\quad =4\pi^2\frac{\alpha_{\rm{EM}}}{m_V^2}\,a_V^2,
\label{Eq:SumRuleV2}
\end{align}
where we expressed the helicity cross sections in Eq.(\ref{Eq:SumRuleV1}) in terms of the $\gamma V\to \gamma V$ helicity amplitudes $M_{\lambda_\gamma' \lambda_V',\lambda_\gamma \lambda_V}$.
The integral in Eq.(\ref{Eq:SumRuleV2}) extends over both bound states and open-flavor states. In principle, all neutral intermediate quarkonium states with even C parity can contribute to this sum rule, see Fig. \ref{fig:lbl-scatt-cut}. These are not only scalar $J^{PC} = 0^{++}$, pseudoscalar $0^{-+}$, axial-vector $1^{++}$, and tensor $2^{++}$ mesons, but also quarkonium states that can carry exotic quantum numbers, such as e.g. $1^{-+}$ and $2^{-+}$. In our analysis, we limit ourselves to the states that correspond to conventional quantum numbers and that are expected to be dominant ones below the $B\bar{B}$ threshold. Furthermore, all of the transitions to pseudoscalar $0^{-+}$ mesons corresponding with magnetic dipole transitions (M1) turn out to be strongly suppressed for the bottomonium transitions, as compared to the transitions to $0^{++},1^{++},2^{++},...$, which correspond predominantly to electric dipole transitions (E1), as discussed below. For bottomonia the anomalous magnetic moment is a small quantity and has been estimated as~\cite{Brambilla:2010cs}
\begin{align}
a_V=\frac{2\,\alpha_s(m_b)}{3\pi}\,. 
\end{align}
Using such value, the sum rule of Eq.~(\ref{Eq:SumRuleV2}) should yield almost zero for bottomonia, i.e., SR $\sim 8\,\text{nb}$.

Unitarity allows us to relate the imaginary part of the $\gamma V\to \gamma V$ helicity amplitude $M_{\lambda_\gamma' \lambda_V',\lambda_\gamma \lambda_V}$ 
to the
$X \to \gamma V$ (for $m_X > m_V$) or  
$V \to \gamma X$ (for $m_V > m_X$) transition amplitudes  ${\cal M}_{\Lambda, \lambda_\gamma \lambda_V}$
\begin{align}
\label{Eq:Unitarity}
&2\,\text{Im}\,M_{\lambda_\gamma' \lambda_V',\lambda_\gamma \lambda_V}
=\sum_X\int d\Gamma_X\,(2\pi)^4\,\delta^{4}(q_1+q_2-p_X)\nonumber\\
&\quad \times 
 {\cal M}_{\Lambda, \lambda_\gamma  \lambda_V}(q_1,q_2;p_x)\,{\cal M}^*_{\Lambda, \lambda_\gamma'\lambda_V'}(q_1,q_2;p_x)\,,
\end{align}
where $\Lambda = \lambda_\gamma - \lambda_V$ denotes the helicity of the quarkonium state $X$, with $\lambda_V$ ($\lambda_\gamma$) being the helicity of the vector quarkonium state $V$ (photon). In the narrow-resonance approximation, Eq.~(\ref{Eq:Unitarity}) can be written as
\begin{align}\label{OpticalTheorem}
&\text{Im}\,M_{+-,+-}=\sum_X\pi\, \delta\left(s-m_X^2\right)\, \abs{{\cal M}_{2, +1 -1}}^2, \nonumber\\
&\text{Im}\,M_{++,++}=\sum_X\pi\, \delta\left(s-m_X^2\right)\, \abs{{\cal M}_{0, +1 +1}}^2, 
\end{align}
which allows us to rewrite the sum rule (\ref{Eq:SumRuleV2}) in terms of the helicity-dependent radiative widths $\Gamma_{\Lambda = 0,2}$ for either the $X \to \gamma V$ or $V \to \gamma X$ transitions. For the $X \to \gamma V$ transitions, the helicity radiative widths are given by: 
\begin{align}
    &\Gamma_{0}(X) =  \frac{1}{4\pi}\, \frac{k}{m_X^2}  \frac{1}{2J_X + 1}\,   \abs{{\cal M}_{0, +1 +1}}^2\,, \nonumber \\
    &\Gamma_{2}(X) = \frac{1}{4\pi}\, \frac{k}{m_X^2}   \frac{1}{2J_X + 1}\,  \abs{{\cal M}_{2, +1 -1}}^2\,,\nonumber\\
    &\Gamma^{\rm{EM}}(X)=\sum_{\Lambda} \Gamma_{\Lambda}(X)\,,
    \label{eq:dw-nr} 
\end{align}
where the photon energy is given by  $k = (m_X^2 - m_V^2)/(2 m_X)$ and $\Gamma^{\rm{EM}}(X)$ is the unpolarized radiative width of the corresponding transition. The corresponding expressions for the helicity radiative widths for the $V \to \gamma X$ transitions are obtained from expressions analogous to Eq.~(\ref{eq:dw-nr}) with the replacement $X \leftrightarrow V$.

Substituting Eqs.~(\ref{OpticalTheorem}) and (\ref{eq:dw-nr}) into the sum rule (\ref{Eq:SumRuleV2}) yields the master formula that we will use in this work:
\begin{align} 
\label{eq:sumrule-dw}
&\sum_{X}^{V \to \gamma X } 8\pi^2\,\frac{(2J_V+1)\,  m_V^3}{(m_V^2 - m_X^2)^3} \left(r_{2}(X) - r_{0}(X) \right) \Gamma^{\rm{EM}}(X) \nonumber \\
&+ \sum_{X}^{X \to \gamma V} 8\pi^2\,\frac{(2J_X+1)\, m_X^3}{(m_X^2 - m_V^2)^3}  \left(r_{2}(X) - r_{0}(X) \right) \Gamma^{\rm{EM}}(X) \nonumber \\
&\simeq 0, 
\end{align}
where we neglected the term proportional to $a_V^2$ as well as the continuum contribution from the open-flavor channels. The latter was estimated using the quark-hadron duality argument \cite{Novikov:1977dq} in total analogy to \cite{Pascalutsa:2012pr,Danilkin:2017utg}. It gives a negligible contribution in the range of $\sim 0.3-7.3$ nb, depending on $V=\Upsilon(1S),\Upsilon(2S),\Upsilon(3S)$.
Furthermore, in Eq.~(\ref{eq:sumrule-dw}) we introduced the helicity ratios
\begin{eqnarray}
r_{\Lambda}(X) = \frac{\Gamma_{\Lambda}(X)}{\Gamma^{\rm{EM}}(X)},
\label{eq:ratio}
\end{eqnarray}
which (as will be shown below) are universal constants in the case of E1 transitions, and depend only on the total angular momentum of the state $X$.

\section{Potential model} 
\label{sec:spectr}

Since the relativistic effects in bottomonia are expected to be small, the spectrum and wave functions can be calculated with the help of the Schr\"odinger equation
and a conventional heavy quarkonium potential,
\begin{align}
&H_0\,\psi(\vec{r})=\left[\frac{{\hat{\vec p}}^2}{m_b}+V(r)\right]\psi(\vec{r})=E\,\psi(\vec{r})\,, \nonumber \\
&V(r)= V_0(r)+V_{SD}(r), \nonumber \\ &V_{SD}(r)=V_{SS}(r)+V_{LS}(r)+V_{T}(r), 
\end{align}
where the Cornell potential $V_0$ is the sum of one-gluon exchange ($V_V$) and linear scalar confining ($V_S$) parts~\cite{Eichten:1978tg},
\begin{align}
V_{0}(r)=V_V(r)+V_S(r)= -\frac{4}{3}\frac{\alpha_S}{r} + b\,r ,
\end{align}
while $V_{SD}$ is a spin-dependent part which splits into the spin-spin ($V_{SS}$), spin-orbit ($V_{LS}$), and tensor ($V_T$) potentials. Up to order $1/m_b^2$,  they are given by
\begin{align} \label{eq:nrqm-ss}
&V_{SS}(r)= \frac{32\,\pi\,\alpha_S}{9\,m_b^2}\,\delta(\vec{r})\, \vec{s}_1 \cdot \vec{s}_{2}, \nonumber \\
&V_{LS}(r)= \frac{1}{2\,m_b^2\,r} \left(3\,\dv{V_V}{r} - \dv{V_S}{r} \right) \vec{L} \cdot \vec{S}, \nonumber \\
&V_T(r)= \frac{1}{12\,m_b^2} \left( \frac{1}{r} \dv{V_V}{r} - \dv[2]{V_V}{r} \right) H_T\,, \nonumber \\
&H_T \equiv 6\,\frac{(\vec{S} \cdot \vec{r})^2}{r^2} - 2\vec{S}^2\,,
\end{align}
where $\vec L$ is the relative orbital momentum operator and $\vec{S}=\vec{s}_1+\vec{s}_2$ is the total spin operator of the quark-antiquark system. Typically, the spin-dependent terms are treated using the leading-order perturbation theory. However, in the present work we follow Deng \textit{et al.}~\cite{Deng:2016stx,Deng:2016ktl} and account for them  nonperturbatively. In order to do that, several modifications are needed. 

First of all, one needs to take the matrix elements over operators in the $|L,S,J,j\rangle$ basis (where $j$ is the spin projection of $J$ on a fixed axis)
\begin{eqnarray}
\langle \vec{s}_1 \cdot \vec{s}_{2} \rangle &=& \frac{1}{2}\,S(S+1)-\frac{3}{4}, \nonumber \\
\langle \vec{L} \cdot \vec{S} \rangle &=& \frac{1}{2}\,\left(J(J+1)-L(L+1)-S(S+1)\right),\nonumber\\
\langle H_T  \rangle &=& \frac{4\,\langle \vec{S}^2\vec{L}^2-\frac{3}{2}\,\vec{L} \cdot \vec{S}-3\,(\vec{L} \cdot \vec{S})^2 \rangle}{(2L+1)\,(2L-1)},
\end{eqnarray}
and regularize a $1/r^3$ behavior in the terms $V_{LS}$ and $V_T$ when $r \rightarrow 0$. The most obvious way to do this is to saturate these potentials at some low-distance scale $r_C$, i.e., set $V_{T}(r)=V_{T}(r_c)$ and $V_{LC}(r)=V_{LV}(r_c)$ when $r < r_C$ \cite{Deng:2016ktl}. 

Second, the physical hyperfine interaction corresponds to a smeared $\delta$ function 
\cite{Barnes:2005pb},
\begin{equation} \label{eq:nrqm-ss}
\delta(\vec{r}) \to \left(\frac{\sigma}{\sqrt{\pi}}\right)^3 e^{-\sigma^2 r^2}, \nonumber
\end{equation}
where $1/\sigma$ is a radius of order $\propto 1/m_b$. 

Finally, in order to effectively account for the creation of virtual light $q\bar{q}$ pairs in the Wilson loop, one considers a screening of the confining potential at large distances $r\gg 1/\mu$:
\begin{equation}
b\,r \to \frac{b}{\mu}\left(1-e^{-\mu\,r}\right)\,.
\end{equation}
The unknown parameters were determined in Ref. \cite{Deng:2016ktl} by fitting the spectrum. A fairly good description of the energy levels was achieved with the following choice of parameters: 
\begin{eqnarray}
\alpha_S &=& 0.368, \nonumber \\
m_b &=& 4.757 \; \mathrm{GeV}, \nonumber \\
b &=& 0.206 \; \mathrm{GeV}^2, \nonumber \\
\sigma &=& 3.10 \; \mathrm{GeV}, \nonumber \\
\mu &=& 0.056 \; \mathrm{GeV}, \nonumber \\ 
r_C &=& 0.060 \; \mathrm{fm}.
\end{eqnarray}
In our work we were able to reproduce the results of Ref. \cite{Deng:2016ktl} to an accuracy of less than $1$ MeV in the energy levels and use these results for the masses of the yet unmeasured $\chi_{b0}(3P)$ and $\chi_{b2}(3P)$ states, as well as to evaluate the radiative transition matrix elements.

\section{E1 radiative transitions} 
\label{sec:rad-trans}

The helicity amplitudes entering Eq.~(\ref{eq:dw-nr}) can be expressed, choosing the Coulomb gauge, as
\begin{align}
&{\cal M}_{\Lambda, \lambda_\gamma \lambda_V} \equiv \sqrt{2E_i}\sqrt{2E_f}\,{\cal M}_{f i} \\
&{\cal M}_{f i} 
= \int \dd^3 {\vec x} \; e^{- i \vec k \cdot \vec x}\, \langle \psi_f | \; \vec \varepsilon^\ast_{\lambda_\gamma} \cdot \vec J(\vec x) \; | \psi_i \rangle, 
\label{eq:mel-nr} 
\end{align}
with the definition $\Lambda \equiv \lambda_\gamma - \lambda_V$, where $\vec J(\vec x)$ is the electromagnetic current operator, and where we introduced the labels $i$ (initial) and $f$ (final) instead of $X$ and $V$ to keep further calculations independent of the direction of the transition. Here the initial and final internal states are labeled by $| \psi_i \rangle = | n_i L_i S_i J_i j_i  \rangle$ and $| \psi_f \rangle = | n_f L_f S_f J_f j_f  \rangle$, where $j_i, j_f$ are the spin projections on a fixed axis. By going to the rest frame of the decaying state, we can orient the quantization axis along the photon momentum direction $\vec k$ and identify the spin projections in terms of helicities, e.g., for the $X \to \gamma V$ transitions, as
$j_i = \Lambda$, $j_f = - \lambda_V$. Note that in the matrix element of Eq.~(\ref{eq:mel-nr}) we use the covariant normalization for the electromagnetic field, but initial and final quarkonium states are normalized nonrelativistically.

A multipole expansion of the electromagnetic field allows to express the matrix element of Eq.~(\ref{eq:mel-nr}) as
\begin{eqnarray}
 && {\cal M}_{fi} 
   = - \sqrt{2 \pi} \sum_{J = 1}^\infty (-i)^J
\int \dd^3 {\vec x} \nonumber \\ 
&& \hspace{1.cm}\times\langle \psi_f |  
i \left[\sqrt{J+1} j_{J-1}(k|\vec x|) \vec Y^{-\lambda_\gamma}_{J-1 J} (\hat x) \right.   \nonumber \\
&&\left. \hspace{2.cm}- \sqrt{J} j_{J+1}(k |\vec x|) \vec Y^{-\lambda_\gamma}_{J+1 J} (\hat x) \right]
 \cdot \vec J(\vec x) \;  
\nonumber \\
&&\hspace{1.75cm}
+ \sqrt{2 J + 1}\lambda_\gamma 
 j_J(k |\vec x|) \vec Y^{-\lambda_\gamma}_{J J}(\hat x)  \cdot \vec J(\vec x) \;  | \psi_i \rangle, \nonumber \\
\label{eq:mult}
\end{eqnarray}
where $j_L(k|\vec x|)$ denote the spherical Bessel functions, and $\vec Y^{M_J}_{LJ}(\hat x)$ is the vector spherical harmonic function, defined in terms of the product of spherical harmonics $Y_{L M_L}(\hat x)$ 
with the photon polarization vector $\vec \varepsilon_\lambda$ as
\begin{eqnarray}
\vec Y^{M_J}_{LJ}(\hat x) = \sum_{M_L} \sum_\lambda \langle L M_L, 1 \lambda | J M_J \rangle \, Y_{L M_L}(\hat x)\, \vec \varepsilon_\lambda.
\end{eqnarray}
Furthermore, in Eq.~(\ref{eq:mult}) the first term, proportional to the squared bracket, corresponds to the electric multipole transitions, whereas the last term corresponds to the magnetic multipole transitions.

The long-wavelength limit is governed by $k r \ll 1$, where $r = |\vec x|$ is the size of the bottomonium system. Because $j_L(k r) \sim (k r)^L$ in this limit, the helicity amplitude in Eq.~(\ref{eq:mult}) will be dominated by the lowest multipoles. Out of these multipoles, the dominant one corresponds to the electric dipole (E1) radiative transitions, i.e., for $J = 1$, due to the behavior of the $j_0(kr)$ term. The magnetic dipole transition (M1) as well as higher multipoles are all suppressed by extra powers of $(k r)$. 
For the sum rule tests for the bottomonium system considered in this paper, where we envisage a numerical precision in the $5 - 10 \%$ range, we can therefore use the nonrelativistic electromagnetic current operator $\vec J(\vec x)$ as
\begin{align}
\vec J(\vec x) = e_Q\, \delta^3\left(\vec x - \frac{\vec r}{2}\right) \frac{{\hat{\vec p}}_Q}{m_Q} 
+ e_{\bar Q}\, \delta^3\left(\vec x + \frac{\vec r}{2}\right)  \frac{{\hat{\vec p}}_{\bar Q}}{m_{\bar Q}}, 
\label{eq:current}
\end{align}
where $m_Q = m_{\bar Q} = m_b$, $e_Q = - e_{\bar Q} = -1/3\,e$ is the bottom quark charge, $\vec r$ is the relative vector between quark and antiquark positions, 
and ${\hat{\vec p}}_Q$ and ${\hat{\vec p}}_{\bar Q}$ are the momentum operators for the quark and anti-quark, respectively. This yields the $E1$ matrix element from Eq.~(\ref{eq:mult}):
\begin{align}
&{\cal M}_{f i}^{\mathrm{E1}} 
= - \sqrt{2 \pi}\, e_Q \int \dd^3 {\vec r} \;
 \psi^\ast_f(\vec r)  
\left[\sqrt{2}\, j_{0}\left(\frac{kr}{2}\right) \vec Y^{-\lambda_\gamma}_{0 1} (\hat r) \right. 
\nonumber \\ 
&\left. \hspace{2cm} -  j_{2}\left(\frac{kr}{2}\right) \vec Y^{-\lambda_\gamma}_{2 1} (\hat r) \right]
 \cdot \frac{\hat{\vec p}}{m_Q/2} \;  \psi_i(\vec r), \, 
\label{eq:e1qqbar}
\end{align}
where $\hat{\vec p} = ({\hat{\vec p}}_Q - {\hat{\vec p}}_{\bar Q})/2$ is the relative momentum operator in the quarkonium system.

Furthermore, to describe the E1 radiative transitions for the bottomonium states with $n_i = n_f$ (and to a lesser extent transitions of nearby $n_i$ and $n_f$), the long-wavelength limit  provides a rather good approximation. As will be shown below, these transitions give the dominant contributions to the sum rule. In this limit the matrix element of Eq.~(\ref{eq:e1qqbar}) is given by
\begin{eqnarray}
{\cal M}_{f i}^{\mathrm{E1}} \label{eq:e1-lw}
   &\simeq& - e_Q\,\langle \psi_f | \;  \frac{ \hat{\vec p} }{m_Q/2}  \; |\psi_i \rangle 
   \cdot \vec \varepsilon_{- \lambda_\gamma} \nonumber \\
 &=& - e_Q\,\frac{d}{dt}\langle \psi_f | \;  \vec r  \; |\psi_i \rangle 
   \cdot \vec \varepsilon_{- \lambda_\gamma}
  \nonumber \\
 &=& 
 - i\,e_Q\,\langle \psi_f | \;  [H_0, \vec{r}] \; |\psi_i \rangle \cdot \vec \varepsilon_{- \lambda_\gamma} \nonumber \\
&=&i\,e_Q\,(m_i - m_f)  \langle \psi_f | \; 
\vec{r} \cdot \vec \varepsilon_{- \lambda_\gamma}  \; |\psi_i \rangle\,,
\end{eqnarray}
where we used the Ehrenfest theorem in the transition from the first to the second line. In result, the nonrelativistic (NRel) expression for the E1 radiative transition width is given by
\begin{eqnarray}\label{eq:Ge1}
&&\Gamma^{\mathrm{E1}}_{\mathrm{NRel}} (i \to \gamma f) =
 \frac{e_Q^2}{4\pi} k^3 
 \left( \frac{2 E_f}{m_i} \right)   \\
 &&\hspace{0.5cm}\times \frac{1}{2 J_i + 1} \sum_{j_i} \sum_{j_f} \sum_{\lambda_\gamma \pm 1}
\abs{ \langle \psi_f | \; 
\vec{r} \cdot \vec \varepsilon_{- \lambda_\gamma}  \; |\psi_i \rangle }^2,\nonumber
\end{eqnarray}
where the mass difference $m_i - m_f$ was approximated by the photon energy $k$. 

As we will only consider transitions from $L_i = 1 \to L_f = 0$ or $L_i = 0 \to L_f = 1$ spin-triplet states ($S_i = S_f = 1$) in the following, Eq.~(\ref{eq:Ge1}) reduces to:
\begin{align}\label{eq:GammaE1}
& \Gamma^{\mathrm{E1}}_{\mathrm{NRel}} (\chi_{bJ}(n_i P) \to \gamma  \, \Upsilon(n_f S)) \nonumber \\
&= \frac{e_Q^2}{4\,\pi} k^3 \frac{4}{9}  \left( \frac{E_f}{m_i} \right) 
\abs{\int_0^\infty dr\, r^3\, R_f(r)\, R_i(r)}^2,  
\end{align}
where $R_i$ and $R_f$ denote the radial wave functions of initial and final states. Note that, for the actual calculations, we account for finite-size corrections to the nonrelativistic E1 result by the replacement~\cite{Novikov:1977dq}
\begin{eqnarray}
\langle \psi_f |\, r\, | \psi_i \rangle \to 
\frac{6}{k}\, \langle \psi_f |\, j_1\left(\frac{kr}{2}\right)\, | \psi_i \rangle\,,
\end{eqnarray}
which are practically only important for the subdominant contributions to the sum rule.

A more systematic inclusion of relativistic effects in calculating the E1 decay widths of heavy quarkonia requires estimating the relativistic corrections to the wave functions, in addition to the recoil and finite-size effects. Such early estimates of the relativistic corrections to the heavy quarkonium E1 decay rates in an expansion up to order $v^2/c^2$ were performed using different potential models, including a Richardson type of potential~\cite{Moxhay:1983vu}, a Coulomb-plus-linear Cornell type potential for $r \gtrsim 0.1$~fm, modified to saturate for $r \lesssim 0.1$~fm~\cite{McClary:1983xw}, and a Buchm\"uller-Tye potential with a scalar confining part~\cite{Grotch:1984gf}. To compare the first-order relativistic corrections between different approaches, we express the relativistic (Rel) calculations of the E1 decay widths of bottomonia as
\begin{eqnarray}
\Gamma^{\mathrm{E1}}_{\mathrm{Rel}} (i \to \gamma f) \equiv 
\Gamma^{\mathrm{E1}}_{\mathrm{NRel}} (i \to \gamma f)  \;(1 + \delta).
\label{eq:rel}
\end{eqnarray}
In Table~\ref{tab:relcorr} we compare the relativistic correction factor $\delta$ for the $\chi_{bJ}(1P) \to \gamma \Upsilon(1S)$ and $\chi_{bJ}(2P) \to \gamma \Upsilon(2S)$ decays in the three above-mentioned calculations. For the decays shown in Table~\ref{tab:relcorr}, the bulk of the relativistic corrections comes from the relativistic modifications to the wave functions, and obviously depends on the choice of the potential. For example, for the $\chi_{bJ}(1P) \to \gamma \Upsilon(1S)$ decay rates, Refs.~\cite{Moxhay:1983vu, McClary:1983xw} found corrections of order $\delta \sim +10 \%$, whereas Ref.~\cite{Grotch:1984gf} reported corrections of order $\delta \sim -20 \%$ to $-15  \%$ for the same transitions. 
\begin{table}[h]
\centering
\begin{tabular}{cccc}
\hline\hline
$\delta$ & \cite{Moxhay:1983vu} & \cite{McClary:1983xw} & \cite{Grotch:1984gf} \\ 
\hline%
$\chi_{b0}(1P) \to \gamma \Upsilon(1S)$  & $+0.09$ & $+0.11$ & $-0.14$ \\%
$\chi_{b1}(1P) \to \gamma \Upsilon(1S)$  & $+0.06$ & $+0.11$ & $-0.17$ \\%
$\chi_{b2}(1P) \to \gamma \Upsilon(1S)$  & $+0.05$ & $+0.11$ & $-0.20$ \\%
\hline
$\chi_{b0}(2P) \to \gamma \Upsilon(2S)$  & $+0.04$ & $+0.32$ & $-0.12$ \\%
$\chi_{b1}(2P) \to \gamma \Upsilon(2S)$  & $-0.06$ & $+0.06$ & $-0.24$ \\%
$\chi_{b2}(2P) \to \gamma \Upsilon(2S)$  & $-0.10$ & $-0.11$ & $-0.35$ \\%
\hline\hline
\end{tabular}
\caption{\label{tab:relcorr}%
First-order relativistic correction $\delta$ to the dominant 
$\chi_{bJ}(nP) \to \gamma  \, \Upsilon(nS)$ (for $n = 1,2$) E1 radiative widths, according to 
Eq.~(\ref{eq:rel}), in different approaches.}
\end{table}

In view of the expected corrections in the $10 - 20~\%$ range for the lower bottomonium states, our strategy in minimizing the model uncertainties in the sum rule estimates is to use the experimental values of the E1 decay widths wherever possible. The latter are available for the $\Upsilon(2S) \to \gamma \chi_{bJ} (1P)$, $\Upsilon(3S) \to \gamma \chi_{bJ} (2P)$, and $\Upsilon(3S) \to \gamma \chi_{bJ} (1P)$ transitions. For the $\chi_{bJ} (nP) \to \gamma \Upsilon(nS)$ transitions (for $n = 1,2,3$), for which the absolute E1 decay widths are not known empirically at present, we will compare their calculated values between five different realistic models that are fit to the spectrum. The spread in the $\Gamma^{\rm{E1}}$ model predictions will be taken as an estimate of the error on the E1 decay width. 

To evaluate the sum rule (\ref{eq:sumrule-dw}), besides the unpolarized radiative width, we also need the helicity ratios $r_{\Lambda = 0,2}(X)$. We will work in the E1 approximation, i.e. neglect the M2 transition for $\chi_{b1}(n_i P) \to \Upsilon(n_f S)$ and the M2 and E3 transitions for $\chi_{b2}(n_i P) \to \Upsilon(n_f S)$. In the E1 approximation,  the coefficients $r_{\Lambda}(X)$ can be expressed as ratios of Clebsch-Gordan coefficients and do not depend on the internal structure of the mesons. Their values are shown in Table~\ref{tab:ratios}.

\begin{table}[h]
\centering
\begin{tabular}{ccc}
\hline\hline
        $J_X$ & \quad $r_{0}(X)$\quad& \quad$r_{2}(X)$\\ \hline
        $0$ & 1 & 0 \\
        $1$ & 1/2 & 0 \\
        $2$ & 1/10 & 3/5 \\ \hline\hline
\end{tabular}
\caption{Helicity ratios $r_{\Lambda}(X)$ for $\Lambda = 0,2$, and for $J_X = 0,1,2$.} 
\label{tab:ratios}
\end{table}

Note that in the extreme nonrelativistic limit where the fine structure is neglected, i.e. when the three $\chi_{bJ}(nP)$ states (for $J = 0,1,2$) are degenerate, the corresponding helicity radiative widths all become proportional to the same E1 squared matrix element. As a consequence, the helicity-0 and helicity-2 sum rule contributions of the three $\chi_{bJ}(nP)$ states are given by
\begin{eqnarray}
\sigma_0 &\sim& \Big\{ r_0(\chi_{b0}) + 3 r_0(\chi_{b1}) + 5 r_0(\chi_{b2}) \Big\} \; \Gamma^{\rm{E1}}_{\rm{NRel}}, \nonumber \\
&=&  \left\{ 1 + \frac{3}{2} + \frac{1}{2} \right\} \Gamma^{\rm{E1}}_{\rm{NRel}} = 3 \, \Gamma^{\rm{E1}}_{\rm{NRel}},
\nonumber \\
\sigma_2 &\sim& 5 r_2(\chi_{b2}) \; \Gamma^{\rm{E1}}_{\rm{NRel}}
= 3 \, \Gamma^{\rm{E1}}_{\rm{NRel}}, 
\label{eq:srnrel}
\end{eqnarray}
which shows that in this extreme limit the sum rule holds exactly for the radiative transitions originating from each shell separately. When calculating with realistic potentials below, where the degeneracy between the three $\chi_{bJ}(nP)$ states is lifted, we will nevertheless observe an approximate cancellation between these states. However, the sum rule of Eq.~(\ref{eq:sumrule-dw}) in general only holds when summing over all radiative transitions to or from a given $\Upsilon(nS)$ state.

\section{Results and discussion} \label{sec:sr}

The central objects in the evaluation of the sum rule of Eq.~(\ref{eq:sumrule-dw}) are the radiative transitions of either $X\to V\gamma$ or $V\to \gamma X$. Several of these transitions have been studied by the Crystal Ball, ARGUS, CLEO, BABAR, and Belle collaborations, as summarized by the Particle Data Group (PDG)~\cite{Tanabashi:2018oca}. Absolute radiative widths are known for the transitions of $\Upsilon(mS) \to \gamma \chi_{bJ}(nP)$ states when $m>n$, and are given in Tables~\ref{tab:tab3} 
and \ref{tab:tab4}. For the opposite transitions of $\chi_{bJ}(nP) \to \gamma \Upsilon(mS)$ when $n \geq m$ only branching fractions Br$(\chi_{bJ}(nP)\to \gamma \Upsilon(mS))$ have been measured. In the absence of the total widths these results cannot be converted into partial widths. For these transitions we will use the results of the potential model, based on Ref.~\cite{Deng:2016ktl}, and outlined in Sections \ref{sec:spectr} and \ref{sec:rad-trans}. To account for the theory uncertainty in the E1 radiative transitions, we include the spread between different theory predictions as our error estimate on this quantity. For the nonrelativistic models, we include the predictions from Refs. \cite{Barnes:2005pb,Li:2009nr, Segovia:2016xqb}, which mainly differ in the form of the potential. The result of Ref.~\cite{Li:2009nr} corresponds to the first-order relativistically corrected wave function with a screened potential model. As for the relativized quark model, we refer to Refs. \cite{Barnes:2005pb,Godfrey:2015dia} where the spinless Salpeter equation was solved, and to Ref. \cite{Ebert:2002pp} which relied on the relativized quasipotential approach.

\begin{table}[h]
\centering
\begin{tabular}{cccc}
\hline\hline
$V = \Upsilon(2S)$ & $m_{\chi_{bJ}}$  &  $\Gamma^{E1}_{\text{th}}$ &$\Gamma_{\text{exp}}$ \\%
& [MeV] & [keV] & [keV] \\
\hline%
$V \rightarrow \gamma\,\chi_{b2}(1P)$ & ${9912}$  & ${2.58}_{-0.70}^{+0.00}$& ${2.29}\pm{0.30}$\\%
$V \rightarrow \gamma\,\chi_{b1}(1P)$&${9893}$&${2.28}_{-0.65}^{+0.17}$&${2.21}\pm{0.31}$\\%
$V \rightarrow \gamma\,\chi_{b0}(1P)$&${9859}$&${1.19}_{-0.28}^{+0.43}$&${1.22}\pm{0.23}$\\%
\hline\hline
\end{tabular}
\caption{
Results for bottomonium radiative transitions $\Upsilon(2S) \to \gamma\, \chi_{bJ}(1P)$ in the quark potential model outlined in Sections 3 and 4, compared with experimental values \cite{Tanabashi:2018oca}. For $\Gamma^{E1}_{\text{th}}$ we include the spread between different predictions \cite{Barnes:2005pb,Li:2009nr,Segovia:2016xqb,Godfrey:2015dia,Ebert:2002pp} as our error estimate.}
\label{tab:tab3}
\end{table}

\begin{table}[h]
\centering
\begin{tabular}{cccc}
\hline\hline
$V = \Upsilon(3S)$& $m_{\chi_{bJ}}$ & $\Gamma^{E1}_{\text{th}}$ & $\Gamma_{\text{exp}}$ \\%
& [MeV] & [keV] & [keV] \\
\hline%
$V \rightarrow \gamma\,\chi_{b2}(2P)$&${10269}$&${3.18}_{-0.88}^{+0.00}$&${2.66}\pm{0.57}$\\%
$V \rightarrow \gamma\,\chi_{b1}(2P)$&${10255}$&${2.66}_{-0.75}^{+0.00}$&${2.56}\pm{0.48}$\\%
$V \rightarrow \gamma\,\chi_{b0}(2P)$&${10233}$&${1.31}_{-0.28}^{+0.18}$&${1.20}\pm{0.23}$\\%
\hline%
$V \rightarrow \gamma\,\chi_{b2}(1P)$&${9912}$&${0.20}_{-0.10}^{+1.1}$&${0.20}\pm{0.04}$\\%
$V \rightarrow \gamma\,\chi_{b1}(1P)$&${9893}$&${0.00}_{-0.00}^{+0.16}$&${0.02}\pm{0.01}$\\%
$V \rightarrow \gamma\,\chi_{b0}(1P)$&${9859}$&${0.12}_{-0.11}^{+0.03}$&${0.06}\pm{0.01}$\\%
\hline\hline
\end{tabular}
\caption{
Same as in Table~\ref{tab:tab3} for the radiative transitions $\Upsilon(3S) \to \gamma\, \chi_{bJ}(2P,1P)$. }
\label{tab:tab4}
\end{table}

\begin{figure}[h]
\includegraphics[width=8.5cm]{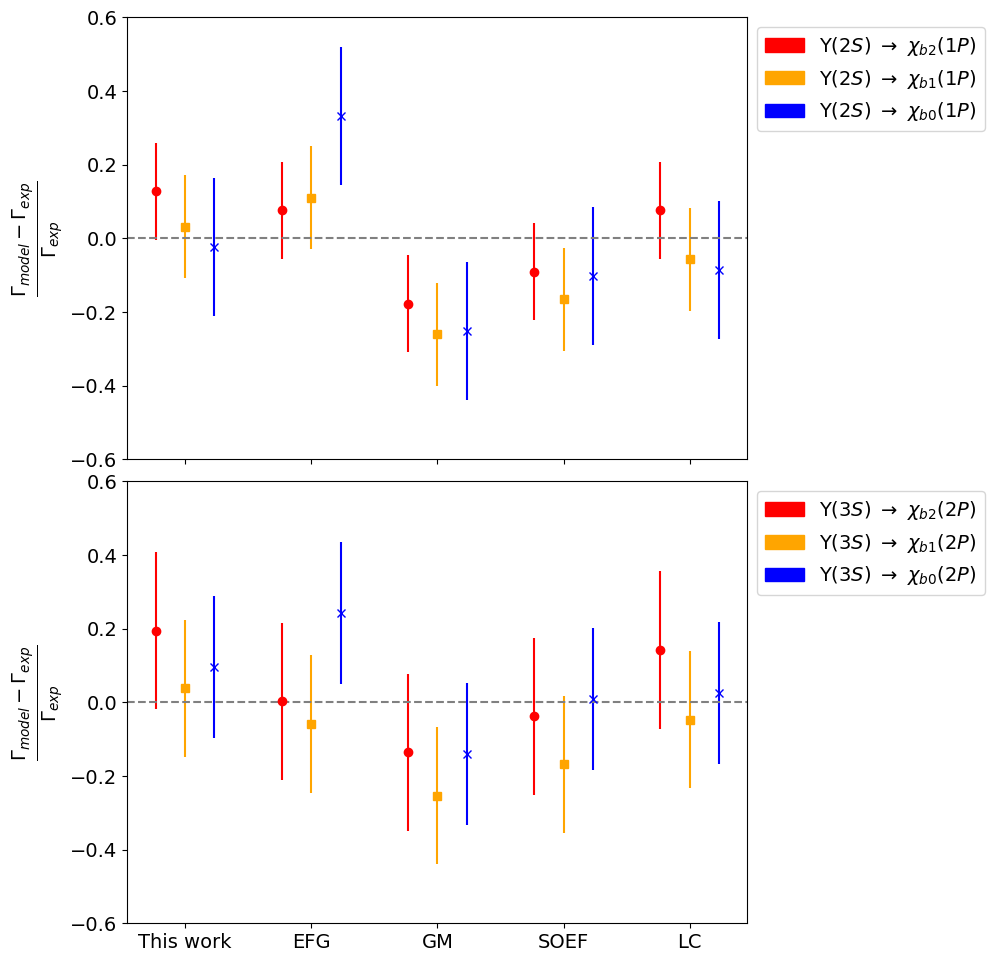}
\caption{Relative comparison between different theoretical and experimental results for the $\Upsilon(2S) \to \gamma\, \chi_{bJ}(1P)$ (upper panel) and 
 $\Upsilon(3S) \to \gamma\, \chi_{bJ}(2P)$ (lower panel) radiative widths. 
Besides the calculation performed here, the models shown are: EFG \cite{Ebert:2002pp}, GM \cite{Godfrey:2015dia}, SOEF  \cite{Segovia:2016xqb}, and LC \cite{Li:2009nr}. }
\label{fig:radwidths_exp}
\end{figure}

We start by comparing the theoretical E1 radiative widths for the $\Upsilon(2S) \to \gamma\, \chi_{bJ}(1P)$ 
and $\Upsilon(3S) \to \gamma\, \chi_{bJ}(2P)$ transitions with their experimental results in Tables~\ref{tab:tab3} and \ref{tab:tab4}, respectively. We see that for nearly all of these transitions the central values (calculated as explained in Sections \ref{sec:spectr} and \ref{sec:rad-trans}) agree with experiment to within 15\%. The spread between the different theoretical values is also in this range, as is illustrated in Fig.~\ref{fig:radwidths_exp}. We therefore feel confident that we can estimate the unknown radiative widths 
$\chi_{bJ}(nP) \to \gamma \Upsilon(mS)$, for $m = 1, 2, 3$,  
with an accuracy at the 20\% level or better.

\begin{table}[t]
\centering
\begin{tabular}{cccc}
\hline\hline
Process & $m_X$ & $\Gamma$ & $SR$ \\%
& [MeV] & [keV] & [$\mu$b] \\%
\hline%
$\chi_{b0}(1P) \rightarrow \gamma\, \Upsilon(1S)$&${9859}$&${24.2}_{-0.4_{\text{th}}}^{+5.7}$&${-1.6}_{-0.4_{\text{th}}}^{+0.0}$\\%
$\chi_{b1}(1P) \rightarrow \gamma\, \Upsilon(1S)$&${9893}$&${30.2}_{-0.7_{\text{th}}}^{+6.4}$&${-2.3}_{-0.5_{\text{th}}}^{+0.1}$\\%
$\chi_{b2}(1P) \rightarrow \gamma\, \Upsilon(1S)$&${9912}$&${36.1}_{-3.5_{\text{th}}}^{+4.1}$&${+4.0}_{-0.4_{\text{th}}}^{+0.5}$\\%
\hline%
\multicolumn{1}{l}{Subtotal}&\multicolumn{3}{r}{${0.2}_{-1.2_{\text{th}}}^{+0.5}$}\\%
\hline%
$\chi_{b0}(2P) \rightarrow \gamma\, \Upsilon(1S)$&${10233}$&${4.4}_{-1.9_{\text{th}}}^{+2.3}$&${-0.04}_{-0.02_{\text{th}}}^{+0.02}$\\%
$\chi_{b1}(2P) \rightarrow \gamma\, \Upsilon(1S)$&${10255}$&${10.7}_{-5.2_{\text{th}}}^{+0.0}$&${-0.14}_{-0.00_{\text{th}}}^{+0.07}$\\%
$\chi_{b2}(2P) \rightarrow \gamma\, \Upsilon(1S)$&${10269}$&${16.9}_{-8.9_{\text{th}}}^{+0.0}$&${+0.35}_{-0.18_{\text{th}}}^{+0.00}$\\%
\hline%
\multicolumn{1}{l}{Subtotal}&\multicolumn{3}{r}{${0.17}_{-0.21_{\text{th}}}^{+0.09}$}\\ 
\hline%
$\chi_{b0}(3P) \rightarrow \gamma\, \Upsilon(1S)$&$10491$&${1.4}_{-1.1_{\text{th}}}^{+0.6}$&${-0.01}_{-0.00_{\text{th}}}^{+0.00}$\\%
$\chi_{b1}(3P) \rightarrow \gamma\, \Upsilon(1S)$&${10512}$&${5.5}_{-4.2_{\text{th}}}^{+0.0}$&${-0.03}_{-0.00_{\text{th}}}^{+0.02}$\\%
$\chi_{b2}(3P) \rightarrow \gamma\, \Upsilon(1S)$&$10528$&${10.7}_{-7.9_{\text{th}}}^{+0.0}$&${+0.10}_{-0.07_{\text{th}}}^{+0.00}$\\%
\hline%
\multicolumn{1}{l}{Subtotal}&\multicolumn{3}{r}{${0.06}_{-0.08_{\text{th}}}^{+0.03}$}\\%
\hline%
\\
\multicolumn{1}{l}{Total}&\multicolumn{3}{r}{${0.4}_{-1.5_{\text{th}}}^{+0.7}$}\\%
\hline%
\hline%
\end{tabular}
\caption{\label{tab:sr-b-scr-yps1s}%
Bottomonium sum rule of Eq.~(\ref{eq:sumrule-dw}) for the radiative transitions involving the $\Upsilon(1S)$ state. Here $m_X$ is the mass of the state $X$, $\Gamma$ is the corresponding radiative decay width to the $\Upsilon(1S)$ state, and $SR$ is the contribution of the corresponding transition to the sum rule.  
The subscript "th" indicates that the theoretical estimate is used. 
}
\end{table}

\begin{table}
\centering
\begin{tabular}{cccc}
\hline\hline
Process & $m_X$ & $\Gamma$ & $SR$ \\
& [MeV] & [keV] & [$\mu$b] \\
\hline
$\Upsilon(2S) \rightarrow \gamma \chi_{b0}(1P)$ & $9859$ 
& $1.22 \pm{0.23}_{\rm{exp}}$ 
& $-3.3$ $\pm {0.6}_{\rm{exp}}$ \\
$\Upsilon(2S) \rightarrow \gamma \chi_{b1}(1P)$ & $9893$ & ${2.21}\pm{0.31}_{\rm{exp}}$&${-5.8}$ $\pm{0.8}_{\rm{exp}}$ \\
$\Upsilon(2S) \rightarrow \gamma \chi_{b2}(1P)$ & ${9912}$ & ${2.29}\pm{0.30}_{\rm{exp}}$ & ${+9.8}$ $\pm{1.3}_{\rm{exp}}$ \\%
\hline%
\multicolumn{1}{l}{Subtotal}&\multicolumn{3}{r}{${0.7}\pm{1.6}_{\text{exp}}$}\\%
\hline%
$\chi_{b0}(2P) \rightarrow \gamma \Upsilon(2S)$&${10233}$&${13.2}_{-2.3_{\text{th}}}^{+0.0}$&${-5.7}_{-0.0_{\text{th}}}^{+1.0}$\\%
$\chi_{b1}(2P) \rightarrow \gamma \Upsilon(2S)$&${10255}$&${15.3}_{-2.0_{\text{th}}}^{+0.6}$&${-7.3}_{-0.3_{\text{th}}}^{+0.9}$\\%
$\chi_{b2}(2P) \rightarrow \gamma \Upsilon(2S)$&${10269}$&${16.7}_{-2.5_{\text{th}}}^{+0.8}$&${+11.2}_{-1.7_{\text{th}}}^{+0.6}$\\%
\hline%
\multicolumn{1}{l}{Subtotal}&\multicolumn{3}{r}{${-1.7}_{-2.0_{\text{th}}}^{+2.5}$}\\%
\hline%
$\chi_{b0}(3P) \rightarrow \gamma \Upsilon(2S)$&$10491$&${2.2}_{-0.5_{\text{th}}}^{+1.5}$&${-0.09}_{-0.06_{\text{th}}}^{+0.02}$\\%
$\chi_{b1}(3P) \rightarrow \gamma \Upsilon(2S)$&${10512}$&${5.0}_{-1.9_{\text{th}}}^{+0.4}$&${-0.27}_{-0.02_{\text{th}}}^{+0.10}$\\%
$\chi_{b2}(3P) \rightarrow \gamma \Upsilon(2S)$&${10528}$&${7.5}_{-3.0_{\text{th}}}^{+0.0}$&${+0.60}_{-0.24_{\text{th}}}^{+0.00}$\\%
\hline%
\multicolumn{1}{l}{Subtotal}&\multicolumn{3}{r}{${0.25}_{-0.32_{\text{th}}}^{+0.12}$}\\%
\hline%
\\
\multicolumn{1}{l}{Total}&\multicolumn{3}{r}{${-0.8}_{-2.3_{\text{th}}}^{+2.6}\pm{1.6}_{\text{exp}}$}\\%
\hline\hline
\end{tabular}
\caption{
\label{tab:sr-b-scr-yps2s} 
Bottomonium sum rule of Eq.~(\ref{eq:sumrule-dw}) for the radiative transitions involving the $\Upsilon(2S)$ state. We took data for the radiative transitions from the PDG~\cite{Tanabashi:2018oca} where available, indicated by the subscript "exp". For the transitions where the absolute radiative widths are not known, we use the predictions based on the model described in this work, indicated by the subscript "th". For the latter, we show the spread in the theoretical calculations as an estimate of the theoretical model error.}
\end{table}

Having compared the theoretical results for the radiative widths of the bound $\Upsilon$ states with available data, we are now in the position to quantitatively verify the sum rule (\ref{eq:sumrule-dw}). In Table~\ref{tab:sr-b-scr-yps1s} we show the sum rule evaluation for the $\Upsilon(1S)$ state. As the absolute radiative widths for the $\chi_{bJ}(nP) \to \gamma \Upsilon(1S)$  transitions are not known, we are using our model estimates to evaluate the sum rule. As theoretical uncertainty in the sum rule evaluation we take the spread among the different models discussed above, and tested on the known radiative widths in Fig.~\ref{fig:radwidths_exp}. From Table~\ref{tab:sr-b-scr-yps1s} we observe a hierarchy of the $\chi_{bJ}(nP) \to \gamma \Upsilon(1S)$ contributions for different shells $n$. As different bottomonia contribute to Eq.~(\ref{eq:sumrule-dw}) with a weighting proportional to the inverse of the third power of the mass difference between the participating bottomonium states, one sees that the individual sum rule contributions from the $1P$ states are more than an order of magnitude larger than those from the $2P$ states, which are about a factor of 4 more important than the $3P$ state contributions. This observed hierarchy also shows that states with $n \geq 4$ are expected to contribute in the few-percent range at most.  One also notices that for each shell ($n = 1, 2, 3$) the sum rule is satisfied well within the theoretical error. For the dominant $n = 1$ transitions, the sum rule result is around 5\% of the dominant  helicity-2 contribution from the $\chi_{b2}(1P)$ state. The cancellation between helicity-0 and helicity-2 contributions for each shell was already discussed following  Eq.~(\ref{eq:srnrel}), being exact in the extreme nonrelativistic limit when the fine structure is neglected. We now see that when using realistic potentials, for which the degeneracy is lifted within each shell, the cancellation is still quite accurate numerically. For the total sum rule, we have an agreement at the 10\% level of the dominant helicity-2 contribution.   

In Tables~\ref{tab:sr-b-scr-yps2s} and \ref{tab:sr-b-scr-yps3s} we show the corresponding results for the $\Upsilon(2S)$ and $\Upsilon(3S)$ states, respectively. For these states we have partial experimental information available on the radiative transitions. We are thus able to also test the sum rule in a more model-independent way when using only experimental data. Besides, for the transitions that are not known experimentally, we are using the theory estimates with their model error range as discussed above. 

For the $\Upsilon(2S)$ state, as shown in Table~\ref{tab:sr-b-scr-yps2s}, the transitions to the $\chi_{bJ}(1P)$ states are all known experimentally to around 15\% precision. The sum rule is seen to hold experimentally for this shell at the 10\% level of the dominant helicity-2 contribution (0.7 $\mu$b vs 9.8 $\mu$b). The same quality of agreement is also found for the second dominant shell in this case, $n = 2$, based on the theoretical estimates for the $\chi_{bJ}(2P) \to \gamma \Upsilon(2S)$ transitions. When evaluating the sum rule for the first three shells, one finds an agreement of better than 5\% of the total helicity-2 contribution, concluding again that this sum rule is well satisfied within the theoretical and experimental error estimates.

\begin{table*}[ht]
\renewcommand*{\arraystretch}{1.4}
\centering
\begin{tabular}{cccc}
\hline\hline
Process & $m_X$ & $\Gamma$ & $SR$ \\%
& [MeV] & [KeV] & [$\mu$b] \\
\hline%
$\Upsilon(3S) \rightarrow \gamma \chi_{b0}(1P)$&${9859.44}$&${0.055}\pm{0.013}_{\text{exp}}$&${-0.0056}\pm{0.0013}_{\text{exp}}$\\%
$\Upsilon(3S) \rightarrow \gamma \chi_{b1}(1P)$&${9892.78}$&${0.018}\pm{0.012}_{\text{exp}}$&${-0.0011}\pm{0.0007}_{\text{exp}}$\\%
$\Upsilon(3S) \rightarrow \gamma \chi_{b2}(1P)$&${9912.21}$&${0.201}\pm{0.043}_{\text{exp}}$&${+0.0142}\pm{0.0030}_{\text{exp}}$\\%
\hline%
\multicolumn{1}{l}{Subtotal}&\multicolumn{3}{r}{${0.0075}\pm{0.0034}_{\text{exp}}$}\\%
\hline%
$\Upsilon(3S) \rightarrow \gamma \chi_{b0}(2P)$&${10232.50}$&${1.20}\pm{0.23}_{\text{exp}}$&${-7.6}\pm{1.5}_{\text{exp}}$\\%
$\Upsilon(3S) \rightarrow \gamma \chi_{b1}(2P)$&${10255.46}$&${2.56}\pm{0.48}_{\text{exp}}$&${-15.1}\pm{2.8}_{\text{exp}}$\\%
$\Upsilon(3S) \rightarrow \gamma \chi_{b2}(2P)$&${10268.65}$&${2.66}\pm{0.57}_{\text{exp}}$&${+24.0}\pm{5.1}_{\text{exp}}$\\%
\hline%
\multicolumn{1}{l}{Subtotal}&\multicolumn{3}{r}{${1.3}\pm{6.0}_{\text{exp}}$}\,\quad\\%
\hline%
$\chi_{b0}(3P) \rightarrow \gamma \Upsilon(3S)$&${10491.40}$&${7.6}_{-0.7_{\text{th}}}^{+0.9}$&${-11.8}_{-1.4_{\text{th}}}^{+1.1}$\\%
$\chi_{b1}(3P) \rightarrow \gamma \Upsilon(3S)$&${10512.00}$&${9.4}_{-1.0_{\text{th}}}^{+0.6}$&${-14.3}_{-0.9_{\text{th}}}^{+1.5}$\\%
$\chi_{b2}(3P) \rightarrow \gamma \Upsilon(3S)$&${10528.24}$&${11.2}_{-1.9_{\text{th}}}^{+0.0}$&${+21.3}_{-3.6_{\text{th}}}^{+0.0}$\\%
\hline%
\multicolumn{1}{l}{Subtotal}&\multicolumn{3}{r}{${-4.8}_{-5.9_{\text{th}}}^{+2.6}$} \quad \quad\\%
\hline%
\\
\multicolumn{1}{l}{Total}&\multicolumn{3}{r}{${-3.6}_{-5.9_{\text{th}}}^{+2.6}\pm{6.0}_{exp}$}\\%
\hline\hline
\end{tabular}
\caption{\label{tab:sr-b-scr-yps3s}%
Same as Table~\ref{tab:sr-b-scr-yps2s} for the bottomonium sum rule of Eq.~(\ref{eq:sumrule-dw}) for the radiative transitions involving the $\Upsilon(3S)$ state.}
\end{table*}

Similar conclusions can also be reached for the $\Upsilon(3S)$ state, shown in Table~\ref{tab:sr-b-scr-yps3s}. The two dominant shells contributing in this case are $n = 2$ and $n = 3$, with the  $\Upsilon(3S) \to \gamma \chi_{bJ}(2P)$ decays widths being known experimentally, and the $\chi_{bJ}(3P) \to \gamma  \Upsilon(3S)$ being estimated theoretically. One sees for the transitions to the $n = 2$ shell that the sum rule is seen to hold experimentally to around 5\% of the dominant helicity-2 contribution (1.3 $\mu$b vs 24.0 $\mu$b). For the $n = 3$ shell contribution, the theoretical estimate shows this to hold at the 20~\% level. When again evaluating the sum rule for the first three shells, one finds an agreement of better than 10\% of the total helicity-2 contribution, concluding again that this sum rule is well satisfied within the theoretical and experimental error estimates.

\section{Conclusion} 
\label{sec:conclude}

In this work, we generalized a forward light-by-light scattering sum rule to the case of radiative transitions between quarkonium states with defined total helicity, of which one has $J^{PC} = 1^{--}$ quantum numbers. The sum rule requires data on radiative transitions in its evaluation. We tested this sum rule on bottomonium vector states. For the transitions of $\Upsilon(mS) \to \gamma \chi_{bJ}(nP)$ states with $m>n$, for which absolute radiative widths are known, we used those data in the sum rule evaluation. For the transitions of $\chi_{bJ}(nP) \to \gamma \Upsilon(mS)$ when $n \geq m$, for which only branching fractions have been measured, we provided theoretical estimates within a potential model. We  considered the spread between similar  approaches in the literature as an estimate for the model error. We checked the potential model on the known $\Upsilon(2S) \to \gamma \chi_{bJ}(1P)$ and $\Upsilon(3S) \to \gamma \chi_{bJ}(2P)$ transitions and found that the theoretical estimates agree with experiment to within 15\%. We then tested the helicity sum rule for the $\Upsilon(1S)$, $\Upsilon(2S)$, and $\Upsilon(3S)$ states. For all three cases we observed that, due to a cancellation between transitions involving $\chi_{b0}, \chi_{b1}$, and $\chi_{b2}$ states, the sum rule is satisfied within experimental and theoretical error estimates. For the total sum rule, a cancellation at the 5 - 10\% level of the dominant helicity-2 contribution was observed. Furthermore, we also observed that for each shell ($n = 1, 2, 3$) the sum rule is satisfied well within the theoretical error.  Having tested this sum rule for the low-lying bottomonium states, it may now be applied to charmonia, where one expects relativistic corrections to potential model results to be more important. Furthermore, as a next step such a sum rule may be used as a tool to investigate the nature of exotic states in the charmonium and bottomonium spectrum, as the radiative transitions involving exotic states containing heavy quarks are proportional to the overlap of initial and final wave functions. In a future extension to the charmonium sector, we plan to include in our analysis first radiative transitions for exotic states which have already been observed at BABAR, Belle, BESIII, and LHCb:  $X(3872) \to \gamma J/\psi$~\cite{Aubert:2006aj,Bhardwaj:2011dj}, $X(3872) \to  \gamma \psi(2S)$~\cite{Aubert:2008ae,Aaij:2014ala}, and $Y(4260) \to \gamma X(3872)$ transitions~\cite{Ablikim:2013dyn} as well as anticipated systematic measurements of these radiative transitions from Belle II \cite{Kou:2018nap}.

\section*{Acknowledgements}
This work was supported by the Deutsche Forschungsgemeinschaft (DFG, German Research Foundation), in part through the Collaborative Research Center [The Low-Energy Frontier of the Standard Model, Projektnummer 204404729 - SFB 1044], and in part through the Cluster of Excellence [Precision Physics, Fundamental Interactions, and Structure of Matter] (PRISMA$^+$ EXC 2118/1) within the German Excellence Strategy (Project ID 39083149).

\bibliographystyle{apsrevM}
\bibliography{raddec2}

\end{document}